\documentclass[10pt,american,english]{article}
\usepackage[T1]{fontenc}
\usepackage[latin9]{inputenc}
\usepackage[letterpaper]{geometry}
\usepackage{float}
\usepackage{textcomp}
\usepackage{amstext}
\usepackage{amssymb}
\usepackage{amsmath}
\usepackage{graphicx}
\usepackage{setspace}
\usepackage[numbers]{natbib}

\makeatletter

\providecommand{\tabularnewline}{\\}

\usepackage{multicol}
\usepackage{fancyhdr}
\usepackage{titlesec}
\usepackage[flushleft]{threeparttable}
\usepackage{caption}
\usepackage{float}



\titleformat*{\section}{\normalsize\bfseries}
\titleformat*{\subsection}{\normalsize\itshape}
\titleformat*{\subsubsection}{\normalsize\itshape}
\titleformat*{\paragraph}{\normalsize\itshape}
\titleformat*{\subparagraph}{\normalsize\itshape}

\makeatother

\usepackage{babel}
\begin{document}
\begin{center}
IAC-18-A7.2.5.47547
\par\end{center}

\medskip{}

\begin{center}
\textbf{Exploring the Kuiper Belt with Sun-diving Solar Sails}
\par\end{center}

\medskip{}

\begin{center}
{\textbf{Elena Ancona}\textsuperscript{a}, \textbf{Roman Ya. Kezerashvili}\textsuperscript{b,c}, and \textbf{Gregory L. Matloff}\textsuperscript{b}}

\par\end{center}

\begin{center}
\textsuperscript{a}Independent researcher, elena.ancona@gmail.com
\par\end{center}

\begin{center}
\textsuperscript{b}New York City College of
Technology, The City University of New York, Brooklyn, NY, USA, rkezerashvili@citytech.cuny.edu, gmatloff@citytech.cuny.edu 
\par\end{center}

\begin{center}
\textsuperscript{c}The Graduate School and University Center, The City University of New York, NY, USA
\par\end{center}

\medskip{}

\begin{center}
\textbf{Abstract}
\par\end{center}

We discuss a possiblity to survey many Kuiper Belt Objects (KBO) with a
single launch using a few small-scale spacecraft, each equipped with solar sails, which could be unfurled from a single interplanetary bus at the
perihelion of that craft's solar orbit. Each small-scale spacecraft would
carry a scientific payload and would be directed to intersect one or more
KBOs. The proposed scenario is the following: the sails are carried as a
payload to a relatively small heliocentric distance (0.1 - 0.3 AU); once at
the perihelion, the sails are deployed. Besides electromagnetic propulsion
due to the solar radiation, another mechanism could be convenient: thermal
desorption, a physical process of mass loss which can provide additional
thrust as heating liberates atoms, embedded on the surface of a solar sail.
Therefore, the sails experience additional propulsive force due to the
thermal desorption that dramatically increases the distance that sails
travel per year.

\bigskip{}

\textbf{Keywords}: Solar Sail, Deep Space Exploration, Thermal Desorption.

\medskip{}

\begin{multicols}{2}

\medskip{}
\section{Introduction}

The Kuiper Belt is a disc-shaped region extending\ 35-50 AU from
the Sun that is populated by volatile-rich objects including the dwarf
planet Pluto and more than 100,000 bodies larger than 100 kilometers across
and as many as a trillion smaller comets. After the discovery of Pluto in
1930, the next discovery of a Kuiper Belt Object (KBO) was in 1992. The NASA
New Horizons Probe encountered Pluto and its satellites in 2015 and is in
route to a second Kuiper Belt destination \cite{Matloff}. It is possible to
survey many Kuiper Belt Objects (KBOs) using a single launch. Many
small-scale spacecraft, each equipped with solar sails, could be unfurled
from a single interplanetary bus at the perihelion of that craft's solar
orbit. Each small-scale spacecraft would carry a scientific payload and
would be directed to intersect one or more KBOs. The following question
arises: Why should we use solar sail?

\columnbreak

\begin{figure*}[b!]
	
	\begin{minipage}[c]{\textwidth} \centering 
		
		\noindent \begin{center}
			\includegraphics[width=15.0cm]{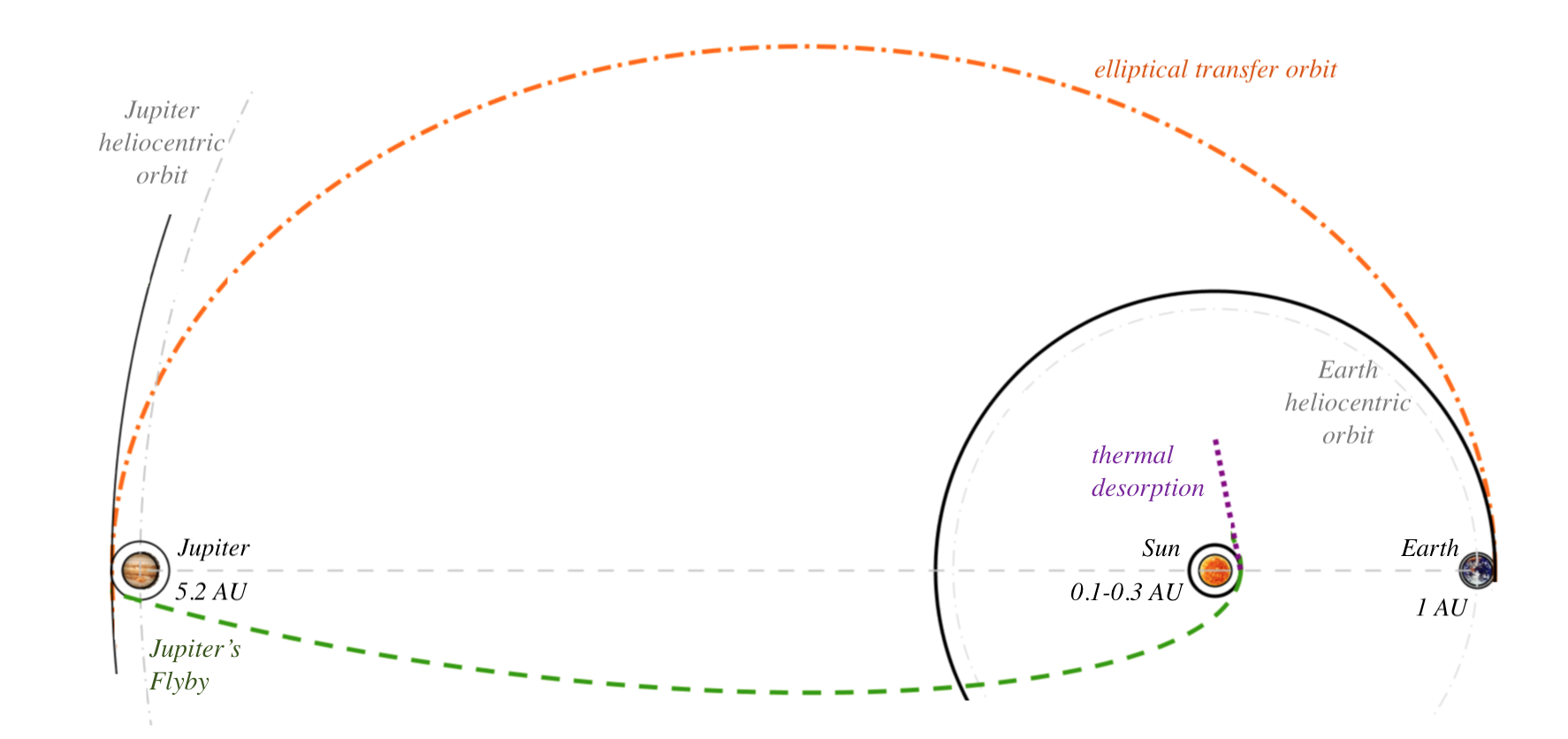}
			\caption{Proposed scenario: elliptical transfer, slingshot and thermal
				desorption (figure not to scale). }
			\label{Fig1_Jupiter_Sun}
			\par\end{center}
		
	\end{minipage}
\end{figure*}

The exhaust speed determines the upper limit for a spacecraft's velocity.
Conventional space transportation systems typically employ chemical or solar
electric propulsion. Chemical propulsion systems have effectively reached
the limit of their maximum performance potential, with further development
activities focused mainly on improving reliability and reducing cost.
Moreover, the chemical rockets are severely limited for such missions
because chemical reactions have a very low energy yield per unit of fuel
mass, and therefore have rather low burn temperatures, less than 1 eV.
Consequently, even though chemical reactions can produce very large thrusts
because of large mass-flow rates for short durations, their exhaust speeds
are necessarily very low. Solar electric propulsion (SEP) is increasingly
being utilized for station keeping, orbit transfer, and deep space
propulsion. Whereas there is still substantial room for improvement in
power, performance, and reliability, SEP systems are not sufficiently
transformative to enable rapid interplanetary transit, let alone missions
into interstellar space.
We would normally expect that the propulsion
methods based on fission, fusion, and antimatter-matter annihilation can
provide high-temperature exhausts because their mass-energy conversion
efficiencies are relatively high: 0.0009, 0.0038, and 1. However, nuclear
electric propulsion \cite{Walker2014} and nuclear thermal propulsion have
the main disadvantage of the large mass of the reactor which necessitates
many heavy lift launches to assemble the transfer vehicle. For space flight,
the main advantage of fusion propulsion would be the very high specific
impulse, and a fusion rocket may produce less radiation than a fission
rocket, reducing the mass needed for shielding. A great deal of work is
required before fusion-propelled space flight can be achieved. The recent
review of fusion concepts for propulsion can be found in Ref. \cite{Cassibry2015}.

Because interstellar distances are very great, a tremendous velocity is
needed to get a spacecraft to its destination in a reasonable amount of
time. Acquiring such a velocity on launch and getting rid of it on arrival
will be a formidable challenge. The solar sail provides very low but
inexhaustible thrust along the direction of motion of a spacecraft bound for
the outer solar system, which increases the distance from the Sun. In this
paper we propose to use a solar sail coated by the material that undergoes
the thermal desorption at a particular temperature \cite{Benford}. Once at
the perihelion of the heliocentric orbit that is determined by the
temperature of desorption of the coating, the solar sail is deployed, the
thermal desorption process can provide an additional significant thrust for
the sail \cite{KezerashviliAA2015}. The sail will be accelerated to the high
velocity due to the thermal desorption of the coating material. After the
desorption coating sublimes away, the sail performs as a conventional solar
sail.

The paper is organized in the following way: in Sec. 2 we describe the sail
craft configuration. The orbital mechanics of a considered scenario is
discussed in Sec. 3. Results of calculations and discussion are given in
Sec. 4, followed by conclusions in Sec. 5.

\section{Sail Spacecraft Configuration}

A solar sail is large sheet of low areal density material whose only source
of energy is the Sun electromagnetic flux. The concept of solar sailing has
been successfully tested by the Japanese IKAROS solar sail spacecraft \cite{IKAROS1, IKAROS2} launched by JAXA and shortly after that by Nanosail-D2
launched by NASA \cite{SailD}. Following these experiences let us consider a
square-shaped sail with area $A$. One important characteristic of a solar
sail is the areal mass density $\sigma $, which is the ratio mass of the
sail $m$\ to the reflective area $A$: $\sigma =\frac{m}{A}$. The total mass
of solar sail spacecraft is composed of the mass of payload $M_{P}$ and mass
of the sail: $\sigma A+M_{P}$. To use the advantage of the acceleration due
to the thermal desorption the reflective area of the sail should be coated
by the material that undergoes the thermal desorption at the particular
temperature \cite{Benford}. We are considering a solar sail coated by
material that undergoes desorption at a particular temperature as a result
of heating by the solar radiation at a particular heliocentric distance \cite{KezerashviliAA2015}. Over the substrate there is a layer of a highly
reflective material, \textit{e. g.} beryllium, aluminum, that is coated by the substance that undergoes the thermal desorption at temperature $T$. \
Let us consider the total mass of the coating $M_{0}$ required for the
desorption. This mass covers the surface area of a solar sail. We assume
that $M_{0}$ is comparable with the mass of the solar sail $\sigma A$, while $M_{c}(t)$ is the coating mass at any instant $t$. We assume that the desorption rate of the coating material $\frac{dM_{c}(t)}{dt}\equiv\frac{dM_{c}}{dt}=-m_{0}$ is
constant. Below we present the analysis for a single solar sail and then
consider using a single launch deployment of a few small-scale spacecraft,
each equipped with solar sails that could be unfurled from a single
interplanetary bus at the perihelion of that craft's solar orbit.

\section{Proposed Scenario}

Let us focus on the orbital dynamics of a solar sail coated with materials
that undergo thermal desorption at a specific temperature, as a result of
heating by solar radiation at the perihelion of a particular heliocentric
orbit. We are considering the following scenario. Using the conventional
spacecraft that carries the solar sails, the transfer occurs from Earth's
orbit to Jupiter's orbit. After that a Jupiter fly-by leads to the
heliocentric orbit with the perihelion close to the Sun, where the
temperature corresponds the temperature of thermal desorption of the coating
material \cite{Ancona}. At this point the sail is deployed and thermal
desorption becomes active. During a short period, the sail is accelerated by
the thermal desorption and as solar radiation pressure. To simplify, during
the process of the thermal desorption one can neglect the acceleration due
to solar radiation because this acceleration is an order of magnitude less
than acceleration due to thermal desorption. It is very important to
remember that in order to have an effective desorption, high temperatures
are required. So one should consider a scenario that takes advantage of a
passage extremely close to the Sun. The schematic diagram for the scenario
is shown in Fig. \ref{Fig1_Jupiter_Sun}. The transfer orbit from Jupiter to the Sun is
the dashed curve, thermal desorption from the perihelion distance
is represented by the dotted curve.
The solid curves represent the orbits of Jupiter and Earth. In this preliminary
analysis the planets are considered point-like. In the heliocentric
reference frame, assuming that Earth's orbit is almost circular, the sail
has to be transferred to an inner orbit closer to the Sun, in order to
escape the Solar System. The transfer between these two coplanar circular
orbits is different and depends on the proposed scenario.

\begin{figure}[H]
	\noindent \begin{centering}
		\includegraphics[width=6.8cm]{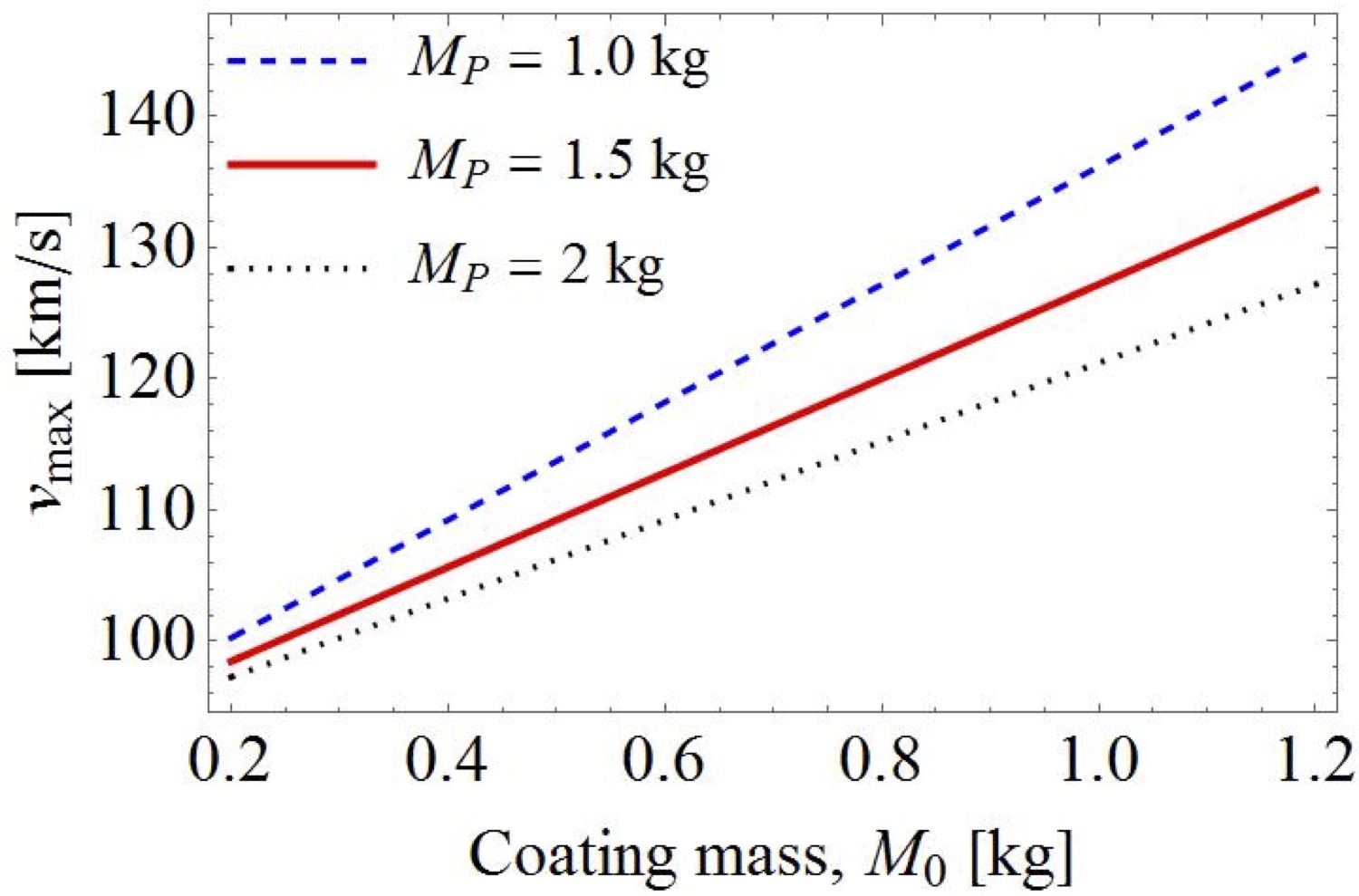}
		\par\end{centering}
	
	\caption{Dependence of maximum velocity $v_{\max }$ on the coating mass $M_{0}$ for different payload masses. Calculation performed for the sail
		areal mass $\protect\sigma = 5$ g/m${^2}$, the desorption process starts at
		the perihelion $r_{p}=0.2$ AU of the heliocentric orbit, the desorption rate 
		$m_{0}=1$g/s and the speed of desorbed atoms is $v_{th}=1236$ m/s. }
	\label{fig:Fig2_MaxVelocity_vs_coating}
\end{figure}

Thermal desorption produces the additional thrust through the release of
atoms of the coating material that have the high exhaust speed. Therefore,
the coating mass at any instant $t$ is $M_{c}(t)\equiv M_{c}=M_{0}-m_{0}t$.
The total time of desorption, that is equal to the time of the acceleration due to the desorption, is $t_{D}=\frac{M_{0}}{m_{0}}$. At the end of the process, when $t=t_{D}$ the coating mass is completely desorbed $M_{c}(t_{D})=0$. The coated unfolded
sail is a moving object with variable mass due to coating material loss
during the desorption period. This fact should be taken into consideration
and one cannot neglect the variation of mass during the acceleration time.
Since the total mass of the coated sail at any instant is $m=\sigma
A+M_{P}+M_{c}$, the force on the sail is:

\begin{equation}
\label{eq:1}
\begin{aligned}
F=\frac{d}{dt}\left( \sigma A+M_{P}+M_{c}\right) v=\\
 =\left( \sigma
A+M_{P}+M_{c}\right) \frac{dv}{dt}+\frac{dM_{c}}{dt}v,
\end{aligned}
\end{equation}

whereas the force due to thermal desorption is:

\begin{equation}
F=\frac{dM_{c}}{dt}v_{th}.
\end{equation}
By comparing these two equations one obtains a differential equation in $v$:

\begin{equation}
\begin{aligned}
\frac{dv}{dt}+\frac{1}{\left( \sigma A+m_{p}+M_{c}\right) }\frac{dM_{c}}{dt}%
v-\\
-\frac{1}{\left( \sigma A+m_{p}+M_{c}\right) }\frac{dM_{c}}{dt}v_{th}=0.
\label{eq:diffeqv}
\end{aligned}
\end{equation}

This result can be rewritten as

\begin{equation}
\frac{dv}{dt}-G(t)v+G(t)v_{th}=0,  \label{RK2}
\end{equation}%
where the time-depended coefficient $G(t)$ is defined as

\begin{equation}
\begin{aligned}
G(t)=\frac{m_{0}}{\left( \sigma A+M_{P}+M_{c}\right) }\equiv\\
\equiv \frac{m_{0}}{%
\left( \sigma A+M_{P}+M_{0}-m_{0}t\right) }.
\label{CoefRK}
\end{aligned}
\end{equation}

The first order differential equation (\ref{RK2}) with time-dependent
coefficient $G(t)$ should be solved assuming that when the mechanism of the
thermal desorption is turned on at $t=0$ the sail's velocity is $v_{p},$ 
\textit{i.e.} $v(0)=v_{p}$, which is the sail's velocity at the perihelion
of the Jupiter's fly-by heliocentric orbit. In Ref. \cite{AnconaKezerashviliArxiv} is given the solution of Eq. (\ref{RK2}) in the
case when one neglects the dependence of the coefficient $G(t)$ on time: $G(t)=$ constant. However, if one considers the time-dependence of the
coefficient $G(t)$, the corresponding solution of Eq. (\ref{RK2}) is

\begin{equation}
v(t)=v_{p}G(t)\frac{\sigma A+M_{P}+M_{c}}{m_{0}}-v_{th}tG(t),
\label{RK_Real}
\end{equation}%
which can be rewritten as:

\begin{equation}
\begin{aligned}
v(t)=\left( \frac{v_{p}(\sigma A+M_{P}+M_{0})}{m_{0}}-v_{th}t\right)\cdot \\
\cdot \frac{m_{0}}{\left( \sigma A+M_{P}+M_{0}-m_{0}t\right) }.  \label{v(t)}
\end{aligned}
\end{equation}

The analysis of Eq. (\ref{v(t)}) shows that the velocity of the sail is
determined by the initial velocity of the sail $v_{p},$ and depends on the
rate of desorption $m_{0}$ and the thermal speed $v_{th}$ of the desorbed
atoms. By the end of the acceleration due to the desorption process at $%
t_{D}=\frac{M_{0}}{m_{0}}$ the sail will gain the maximum velocity

\begin{equation}
v(t_{D})\equiv v_{\max }=v_{p}+\left( v_{p}-v_{th}\right) \frac{M_{0}}{%
\sigma A+M_{P}}.  \label{RKReal_Vmax}
\end{equation}
Eq. (\ref{RKReal_Vmax}) shows that the maximal velocity of the sail is
determined by the initial velocity of the sail $v_{p}$ at the perihelion of
the heliocentric orbit, and ratio of coating mass $M_{0}$ to mass of the
sailcraft $\sigma A+M_{P}$ excluding the coating mass. With this velocity,
the sail continues to accelerate due to the solar radiation at a lower rate
for a longer time interval. The corresponding description of the sail
acceleration due to the solar radiation is well known and can be omitted
here.

\begin{figure}[H]
	\noindent \begin{centering}
		\includegraphics[width=6.5cm]{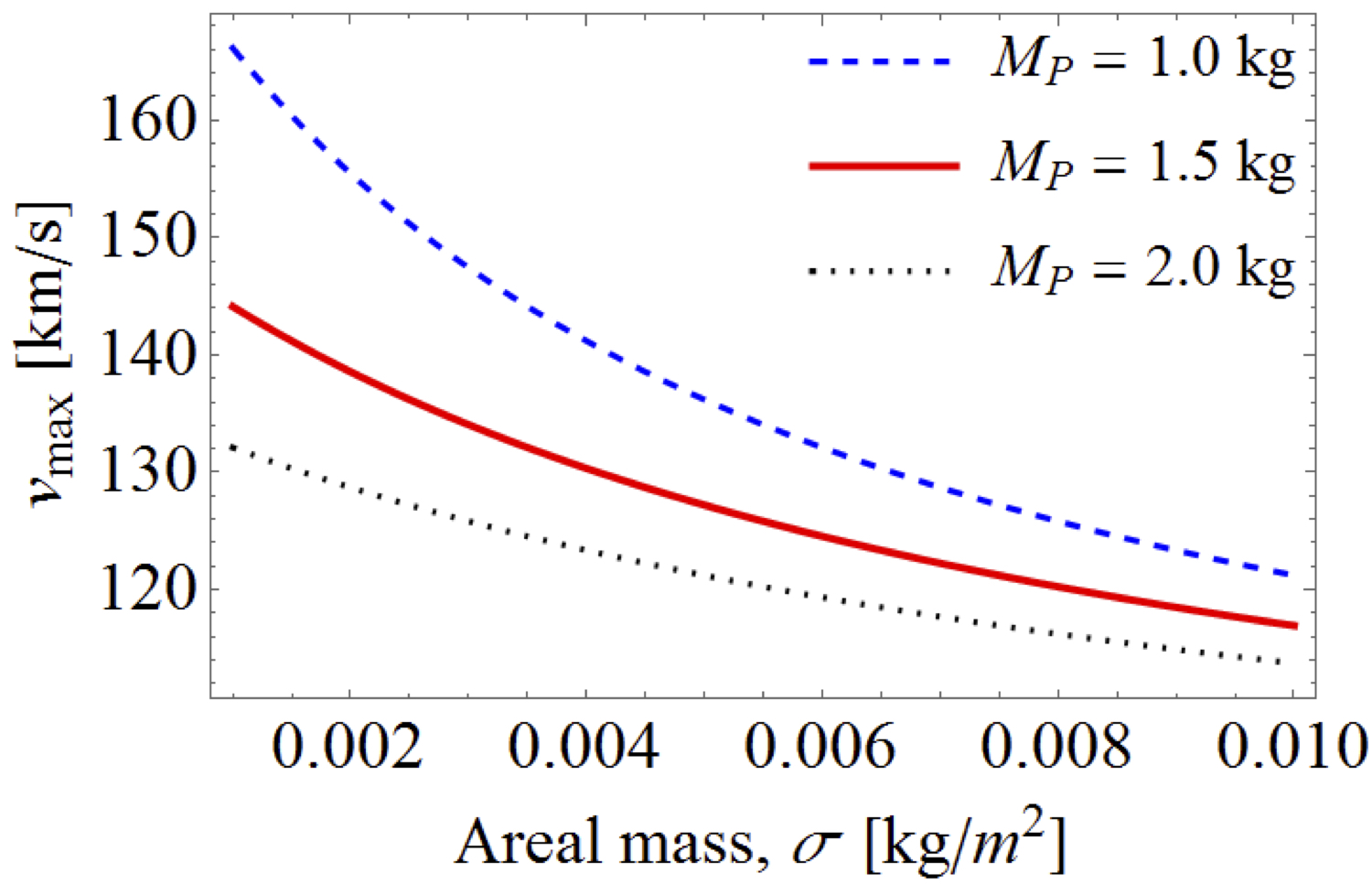}
		\par\end{centering}
	
	\caption{Dependence of maximum velocity $v_{\max }$ on the areal mass $%
		\protect\sigma$ for the different payload masses. Calculation performed for
		the coating mass $M_{0}$ = 1 kg, the desorption process starts at the
		perihelion $r_{P}=0.2$ AU of the heliocentric orbit, the desorption rate $%
		m_{0}=1$ g/s and the speed of desorped atoms is $v_{th}=1236$ m/s.}
	\label{fig:Fig3_Max_Velocity_on_Areal_mass}
\end{figure}

\begin{figure}[H]
	\noindent \begin{centering}
		\includegraphics[width=6.5cm]{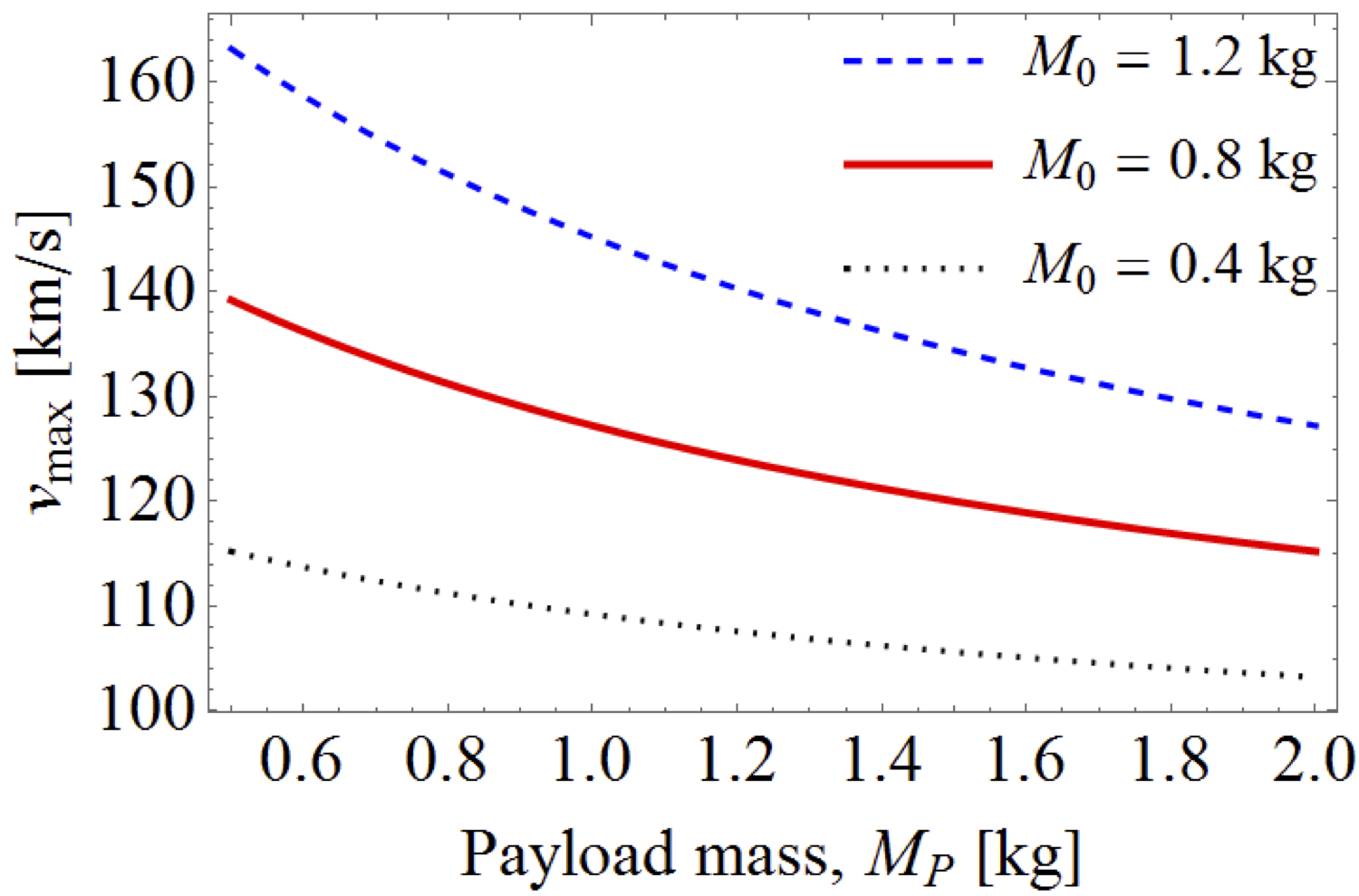}
		\par\end{centering}
	
	\caption{Dependence of maximum velocity $v_{\max }$ on the payload mass $%
		M_{P}$ for the different coating masses. Calculation performed for the sail
		areal mass $\protect\sigma =5$ g/m$^{2}$, the desorption process starts at
		the perihelion $r_{P}=0.2$ AU of the heliocentric orbit, the desorption rate 
		$m_{0}=1$ g/s and the speed of of desorped atoms is $v_{th}=1236$ m/s.}
	\label{fig:Fig4_Max_Velocity_on_payload_mass}
\end{figure}

\section{Results of Calculations and Discussion}

For the sake of brevity we will skip the orbital dynamics calculations,
which are nothing but the application of well-known conventional formulas 
\cite{Bate}. Instead, we go directly to the results that show the positive
contribution of thermal desorption. We are assuming that the elliptical
transfer is performed from Earth's orbit to Jupiter's orbit and the
spacecraft can be in different heliocentric orbits. At the perihelion of the
orbit where the sail should be deployed, the temperature corresponds to the
desorption temperature of the coating material. We present the result for
three perihelions: 0.1 AU, 0.2 AU and 0.3 AU. The corresponding temperature
varies from 736 K at $r_{p}=$0.3 AU to 1149 K at $r_{p}=$0.1 AU \cite{Ancona}%
. The corresponding thermal velocities of carbon atoms vary from 1140 m/s to
1420 m/s. We consider the sail with the following parameters: $A=200$ m$^{2}$%
, $\sigma =5$ g/m$^{2}$. If one deposits the coating mass $M_{0}$ of 0.2 kg
to 1.2 kg for the sail area $A=2\times 10^{2}$ m$^{2}$, the corresponding
thickness of the coating layer varies from about \ $H=0.6$ $\mu $m to $H=3.5$
$\mu $m, assuming that the coating is carbon-based material with the avarage
bulk density 1.74 $\frac{\text{g}}{\text{cm}^{3}}$. We next consider the
dependence of the maximal sail velocity that it gains due to thermal
desorption on the characteristic parameters of the coated solar sail. In
particular, we study the dependence of the maximum velocity that the solar
sail reaches due to the thermal desorption of coating on the deposit coating
mass, $M_{0}$, the areal density of the sail, $\sigma $ and the mass of the
payload, $M_{P}$. The results of calculations are presented in Figs. \ref{fig:Fig2_MaxVelocity_vs_coating}, \ref{fig:Fig3_Max_Velocity_on_Areal_mass} and \ref{fig:Fig4_Max_Velocity_on_payload_mass},
respectively. Figure \ref{fig:Fig2_MaxVelocity_vs_coating} presents the
results for calculations of the dependence of maximal velocity on the
coating mass for different payload masses ($r_{p}=0.2$ AU, $\sigma =5$ g/m$%
^{2})$. One can see that the increase of the coating mass leads to the
increase of maximum velocity, while when the payload mass increases the $%
v_{\max }$ decreases.

\begin{figure}[H]
	\noindent \begin{centering}
	\includegraphics[width=6.5cm]{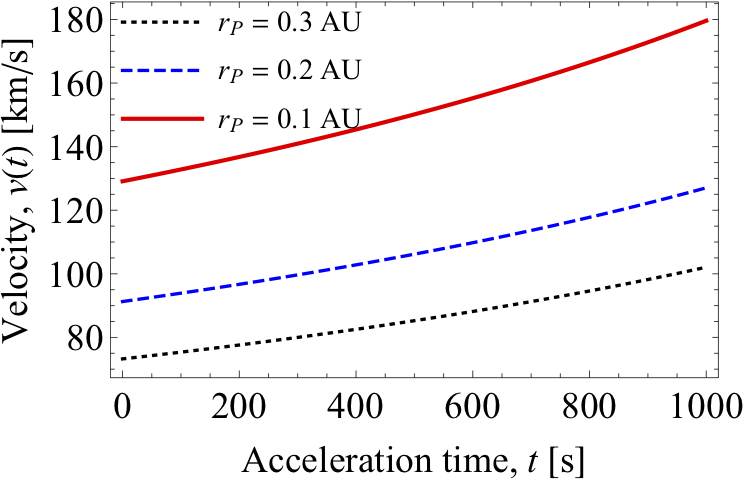}
	\par\end{centering}	
		
\caption{Dependence of velocity of the sail $v(t)$ on time $t$ during the
acceleration due to the thermal desorption when the sail is deployed at the
perihelion of the different heliocentric orbits. Calculation performed for
the sail areal mass $\sigma $ $=5$ g/m$^{2}$, the coating mass $M_{0}=1$.0 kg, the mass of payload $M_{P}=1.5$ kg, the desorption rate $m_{0}=1$ g/s.}
\label{fig:Fig5_Dependence_on_Perihelion}%
\end{figure}

The dependence of the velocity of solar sail during the acceleration process
due to thermal desorption when the sail is deployed at the perihelion of
different heliocentric orbits is presented in Fig. \ref{fig:Fig5_Dependence_on_Perihelion}. These calculations are performed for
the coating mass $M_{0}=1.0$ kg and mass of payload $M_{P}=1.5$ kg. One can
conclude that the closer the solar approach, the higher the velocity that
can be achieved using thermal desorption.

\begin{figure}[H]
	\noindent \begin{centering}
\includegraphics[width=6.5cm]{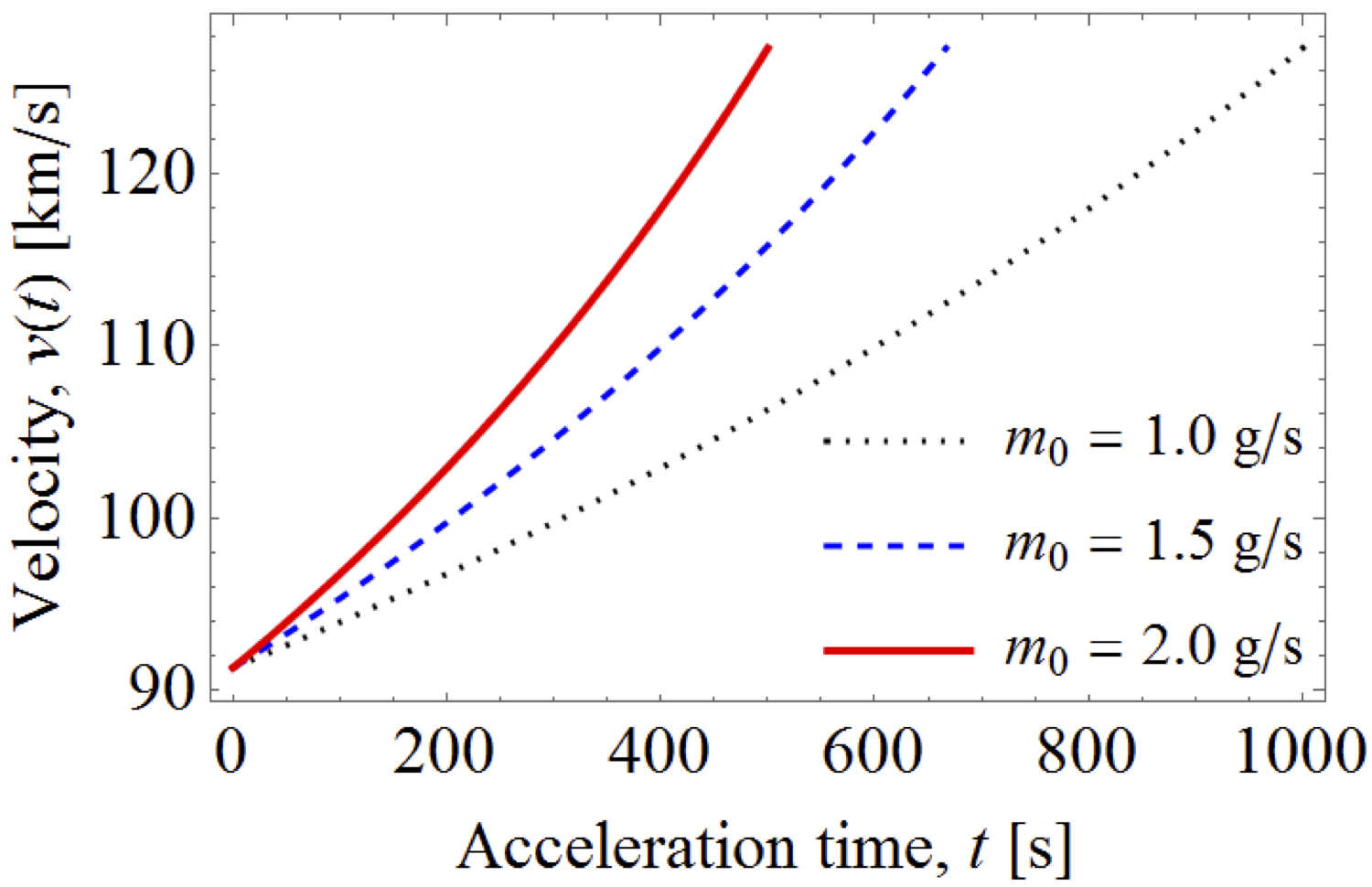}
\par\end{centering}

\caption{Dependence of velocity of the sail $v(t)$ on time $t$ during the
acceleration due to the thermal desorption when the sail is deployed at the
perihelion $r_{P} =0.2$ AU of the different heliocentric orbit. Calculation
performed for the coating mass $M_{0}=1$.0 kg, the desorption rate $m_{0}=1$
g/s and the speed of of desorbed atoms is $v_{th}=1236$ m/s.}
\label{fig:Fig6_Dependence_on_Rate}
\end{figure}

While the maximum velocity of the sail does not depend on the rate of
thermal desorption, the time to reach the maximum velocity depends on the
rate of this process. The time to reach the same maximum velocity decreases
when the rate of the desorption increases. This is clearly illustrated in
Fig. \ref{fig:Fig6_Dependence_on_Rate}, which presents the dependence of the
velocity of the sail on the rate of the desorption. When the desorption rate
increases, the acceleration time decreases and the sail reaches a given $v_{\max}$ in a shorter time.

In Table \ref{Velocities}, we present the results obtained for these three
perihelion values. Calculations are performed for the coating mass $M_{0}=1$
kg required for the desorption and the mass of payload $M_{P}=1.5$ kg. In
Table \ref{Velocities} $v_{P}$ is the sail velocity at perihelion, before
desorption occurs. This velocity is calculated from the orbital dynamics of
a conventional spacecraft for a proposed scenario. By contrast, the final
velocity $v_{max}$ of the sail after the desorption time $t_{D}$ is
calculated by using Eq. (\ref{RKReal_Vmax}), which is the exact solution of
Eq. (\ref{RK2}) with time-depended coefficient $G(t)$. The final cruise
speed $v_{c}$ is obtained from the application of the law of conservation of
energy, and from it comes the distance travelled per year.

\captionof{table}{Scenario for elliptical transfer plus Slingshot plus thermal desorption acceleration. For each perihelion of the heliocentric orbit are presented the corresponding speed before desorption $v_{p}$, the maximum speed $v_{\max }$ from Eq. (\ref{RKReal_Vmax}) after desorption, cruise speed $v_{c}$ and distance $D_{y}$ covered per year. Calculations are performed for the coating mass $M_{0}=1$ kg, desorption rate $m_{0}=1$ g/s and mass of payload $M_{P}=1.5$ kg. }

\begin{singlespace}
	\noindent \begin{center}
		\begin{tabular}{llrrr}
			\hline 
			\multicolumn{5}{c}{Speed and Distance covered}\tabularnewline
			\hline
			
			$r_{P}$ & AU & 0.3 & 0.2 & 0.1\tabularnewline 
			$v_{p}$ & km/s & 73.24 & 91.28 & 129.11\tabularnewline
			$v_{\max }$ & km/s & 102.08 & 127.30 & 180.19\tabularnewline
			$v_{c}$ & km/s & 107.72 & 134.09 & 189.78\tabularnewline
			$D_{y}$ & AU/year & 22.71 & 28.26 & 40.01\tabularnewline
			\hline
			
		\end{tabular}
		\par\end{center}
\end{singlespace}
\label{Velocities}

\section{Conclusions}

Investigation of Figs. 2 - 6 reveals that application of desorption results
in high post-perihelion heliocentric solar sail velocities. Certainly,
application of this technology with state-of-the-art sails can result in
solar-system exit velocities in excess of 100 km/s.

At 100 km/s, or $\sim$20 AU/year, post-perihelion travel times to the
vicinity of Kuiper Belt Objects will be less than 3 years. Factoring in time
required for the Jupiter flyby to perihelion, target KBOs could be
encountered less than 6 years after launch.

The solar-photon sail configuration considered in this paper is conservative
and based on sails that have successfully operated in Low Earth Orbit or
interplanetary space. Recent research reveals that much smaller sails could
be incorporated with highly miniaturized chip-scale spacecraft \cite%
{Weis2014, NASA2015}. It is quite possible that a single dedicated interplanetary
\textquotedblleft bus\textquotedblright\ could deploy many chip-scale sails
at perihelion. Sequential deployment of these sails could be timed to allow
exploration of many small KBOs from a single launch.

\end{multicols}{}
\end{document}